\documentclass[conference]{IEEEtran}
\IEEEoverridecommandlockouts

\usepackage{cite}
\usepackage{amsmath,amssymb,amsfonts}
\usepackage{algorithmic}
\usepackage{graphicx}
\usepackage{textcomp}
\usepackage{xcolor}
\usepackage{float}
\usepackage{hyperref}
\usepackage{booktabs}
\usepackage{multirow}
\usepackage{makecell}

\def\BibTeX{{\rm B\kern-.05em{\sc i\kern-.025em b}\kern-.08em
    T\kern-.1667em\lower.7ex\hbox{E}\kern-.125emX}}
\begin{document}

\title{Reliable Quasi-Static Post-Fall Floor-Occupancy Detection Using Low-Cost Millimetre-Wave Radar
}

\author{\IEEEauthorblockN{Huy Trinh}
\IEEEauthorblockA{
\textit{Dept. Electrical and Computer Engineering} \\
\textit{University of Waterloo}\\
Waterloo, Canada \\
h3trinh@uwaterloo.ca}
\and
\IEEEauthorblockN{Phuong Thai}
\IEEEauthorblockA{
\textit{Dept. Chemistry and Biology} \\
\textit{Toronto Metropolitan University}\\
Toronto, Canada \\
phuong.thai@torontomu.ca}
\and
\IEEEauthorblockN{Elliot Creager}
\IEEEauthorblockA{
\textit{Dept. Electrical and Computer Engineering} \\
\textit{University of Waterloo}\\
Waterloo, Canada \\
creager@uwaterloo.ca}
\and
\IEEEauthorblockN{George Shaker}
\IEEEauthorblockA{
\textit{Dept. Electrical and Computer Engineering} \\
\textit{University of Waterloo}\\
Waterloo, Canada \\
gshaker@uwaterloo.ca}
}

\maketitle
\begin{abstract}
As the population ages rapidly, long-term care (LTC) facilities across North America face growing pressure to monitor residents safely while keeping staff workload manageable. Falls are among the most critical events to monitor due to their timely response requirement, yet frequent false alarms or uncertain detections can overwhelm caregivers and contribute to alarm fatigue. This motivates the design of reliable, whole end-to-end ambient monitoring systems from occupancy and activity awareness to fall and post-fall detection. In this paper, we focus on robust post-fall floor-occupancy detection using an off-the-shelf 60 GHz FMCW radar and evaluate its deployment in a realistic, furnished indoor environment representative of LTC facilities. Post-fall detection is challenging since motion is minimal, and reflections from the floor and surrounding objects can dominate the radar signal return. We compare a vendor-provided digital beamforming (DBF) pipeline against a proposed preprocessing approach based on Capon/minimum variance distortionless response (MVDR) beamforming. A cell-averaging constant false alarm rate (CA-CFAR) detector is applied and evaluated on the resulting range–azimuth maps across 7 participants. The proposed method improves the mean frame-positive rate from 0.823 (DBF) to 0.916 (Proposed). 
\end{abstract}

\section{Introduction}
Fall incidents remain a major cause of injury and loss of independence among older adults, leading to emergency visits and long-term care admissions~\cite{Terroso}. In clinical settings, the term ”long-lie” is often referred to a situation in which an individual has fallen and remains on the floor for one hour or longer~\cite{10.1371/journal.pone.0177510, doi:10.33151/ajp.8.4.86,NYMAN_VICTOR_2014,Kubitza2023ConceptOT}. Long-lies can cause hypothermia, dehydration, pressure ulcers, muscle damage, infections, and even loss of consciousness~\cite{Kubitza2023ConceptOT}. Most of the current fall-detection literatures focus on event detection (classifying a fall at the moment it occurs) using wearable inertial sensors, vision systems, or ambient sensors~\cite{9018226}. A study by Omar et al.~\cite{s25247423} reviews various advanced fall detection technologies for older adults. The practical deployment in real homes and care facilities, however, can remain challenging. Wearables can be forgotten, removed, broken and malfunction after a fall; camera-based approaches raise privacy concerns and can fail under occlusion, lighting changes, or clutter~\cite{9704291}. As a result, these risks pose an urgent need for post-fall detection systems that can promptly and continuously detect floor occupancy and initiate an appropriate response. 

Radio-frequency (RF) based sensing—particularly millimetre-wave (mmWave) radar has emerged as a compelling alternative to vision-based systems and can be used for future in-home monitoring and ambient assisted living (AAL) due to its advantages in privacy preservation, robustness in low-light or occluded settings, and low data bandwidth requirements~\cite{7426551, bios7040055,10554983}. While there has been various research on detecting people falling using biomedical radar systems, there is currently little research on once they have fallen or in the post-fall phase~\cite{10136884, TEWARI2023107315}. Unlike detecting the falling event, the clinical requirement after a fall is frequently not the precise classification of the dynamic event, but the reliable detection of the post-fall state: an individual lying quasi-statically on the floor. This has long been known to be more challenging than detecting moving targets due to the fact that it can hide weak reflections and raise the background floor, therefore hindering precise detection, tracking and localization~\cite{10118759}. Furthermore, multipath and partial occlusions can suppress or spatially distort the returned signal from subjects lying on the floor. And finally, different radar placements, arbitrary after-fall directions and fall positions can pose other challenges in reliable detection of the system~\cite{6345918,Su2018RadarPF, s23115031}. In this work, our main contributions can be summarized in two points:
\begin{itemize}
    \item We formulate quasi-static post-fall floor-occupancy detection as a reliability-critical sensing problem distinct from fall-event classification and set up the experiments in real-world long-term care facility conditions. 
    \item We propose a range-azimuth (RA) preprocessing pipeline based on minimum variance distortionless response (MVDR)/Capon beamforming and compare it against the vendor digital beamforming (DBF) baseline using the same Cell-Averaging Constant False Alarm Rate (CA-CFAR) detection stage post-processing. We quantify both average performance, worst-case analysis, and reliability/coverage across thresholds. 
\end{itemize}
The rest of the paper is organized as follows. 
Section~\ref{sec:methodologies} describes the radar signal processing pipeline, the vendor DBF baseline, our proposed Capon/MVDR preprocessing, and the CA-CFAR detector with its tuning procedure. Section~\ref{sec:results} presents our experiment setup against the vendor's standard beamforming method, reports results for participant-level comparisons and reliability analysis across viewpoints and floor locations. Finally, Section~\ref{sec:conclusion} concludes the study, discusses the implications and future work.

\label{sec:introduction}
\section{Methodologies}
\label{sec:methodologies}
\subsection{Experiment Setup and Data Collection}
\begin{figure}[H]
\centerline{\includegraphics[width=\linewidth]{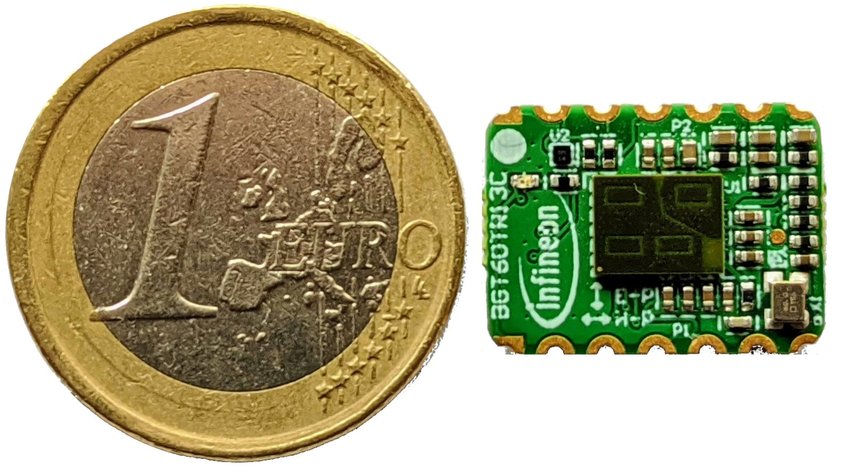}}
\caption{Infineon XENSIV\textsuperscript{\texttrademark} BGT60TR13C Radar. \cite{infineon2024bgt60tr13c} }
\label{fig:bgt60_image}
\end{figure}

\begin{figure}[htb] 
\centering 
\includegraphics[width=\linewidth]{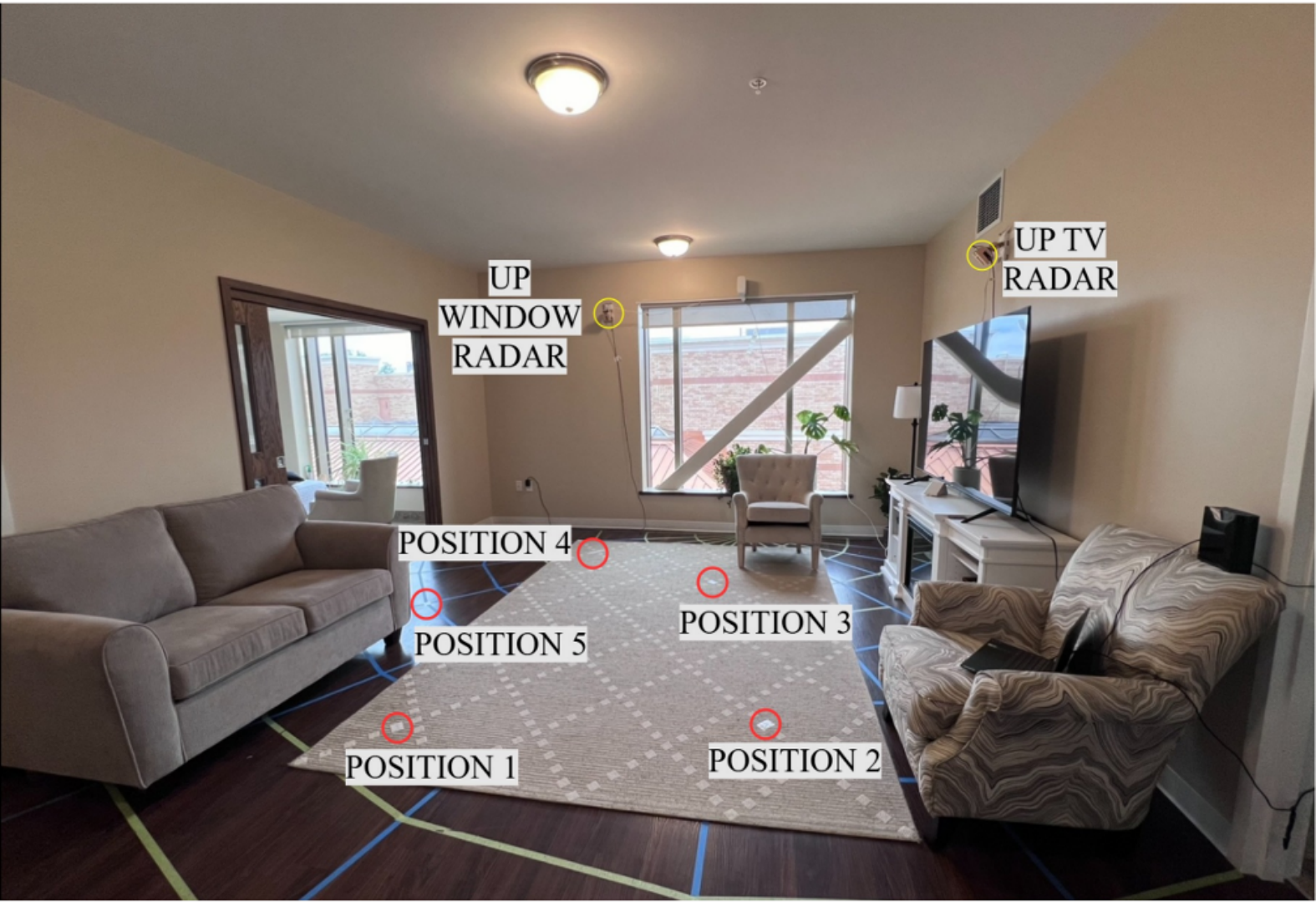}
\caption{Experimental setup and viewpoints. Two wall-mounted radar viewpoints (TV-side and Window-side) are used to observe post-fall floor occupancy across five floor locations (P1--P5) under real room clutter and occlusions.}
\label{fig:experiment_setup}
\end{figure}

\begin{table}[htb]
\centering
\caption{Radar hardware and waveform configuration (BGT60TR13C).}
\label{tab:radar-hw}
\begin{tabular}{l l}
\hline
\textbf{Parameter} & \textbf{Value} \\
\hline
Band / Center frequency & 58–63.5\,GHz (center $\approx$ 60\,GHz) \\
Sweep bandwidth $B$ & $\approx$ 500\,MHz \\
Frame rate & 10\,Hz \\
Chirps per frame $N_{\text{chirp}}$ & 128 \\
Samples per chirp $N_{\text{sample}}$ & 64 \\
Range resolution $\Delta r$ & 0.30\,m \\
Max range  & 9.6\,m \\
Max unambiguous speed  & 3\,m/s \\
\hline
\end{tabular}
\end{table}
\begin{figure*}[!t] 
  \centering
  \includegraphics[
    width=\textwidth,
    keepaspectratio,          
    trim={0 236 0 300pt},        
    clip
  ]{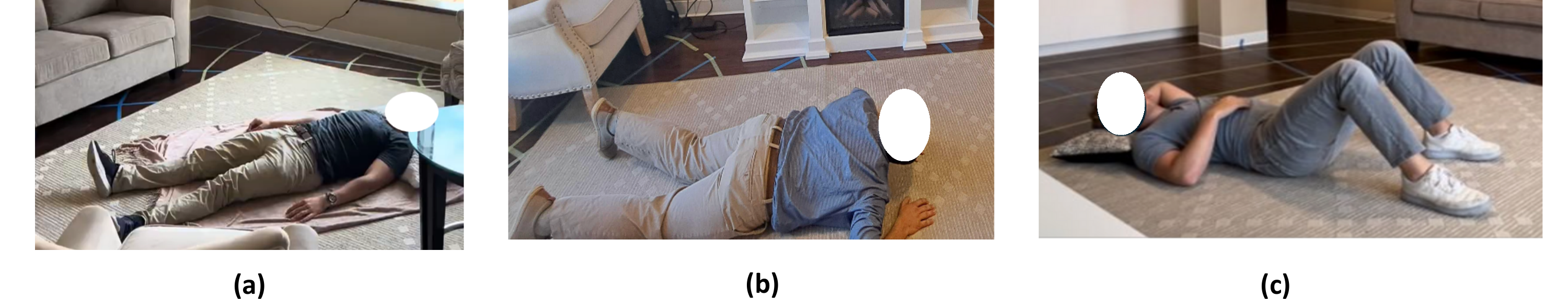}
  \caption{Few example post-fall floor postures from the dataset. Subjects lie on the floor in multiple quasi-static poses (a) supine with arms extended, (b) side-prone, and (c) supine with knees bent. Furniture is intentionally placed near the subject to create multipath and partial occlusions.}
  \label{fig:diferent_falls}
\end{figure*}
For this study, we collected a dedicated dataset of quasi-static post-fall postures in a long-term care (LTC) facility using an off-the-shelf 60\, GHz mmWave Frequency-Modulated Continuous-Wave (FMCW) radar (Infineon XENSIV\textsuperscript{\texttrademark} BGT60TR13C)\cite{infineon2024bgt60tr13c} shown in Fig.~\ref{fig:bgt60_image}. The selected radar has one transmitter antenna (Tx) and three receiver antennas (Rx). With this alignment, it is possible to obtain both the azimuth and elevation angle of arrival from the reflected signals. The radar configuration used in all recordings is summarized in Table~\ref{tab:radar-hw}. With this setting, the radar can offer a maximum detection range of 9.6 meters, which is enough to cover the measuring areas. 
To reflect real indoor assisted-living conditions, data were collected in a furnished room with common clutter such as a sofa bed, armchairs, a coffee table, a standing lamp, shelves, etc. The furniture layout is intentionally kept close to the subject to create realistic multipath and partial occlusions. Two radar viewpoints are used: a TV-side view and a Window-side view. Each sensor is wall-mounted and tilted downward, approximately 2.5\,m above the floor (Fig.~\ref{fig:experiment_setup}).
To mimic post-fall conditions safely, subjects are asked to lie on the floor in several representative postures (e.g., facing up, side-lying, and knees-bent) as shown in Fig.~\ref{fig:diferent_falls}. For each viewpoint, the subject is placed at five floor locations (P1--P5), covering cases near the wall, near furniture edges, and open-floor regions. These positions are annotated as expected ground-truth regions and will be used for evaluation later in Section~\ref{sec:results}.
For each subject, we recorded 2 minutes at 5 distinct positions and orientations relative to the sensors to vary the aspect angle, range, and background layout. Besides, for each measurement session, we placed these furniture randomly to create angular diversity to mimic different room layouts in the LTC. This enables our final datasets to be very diverse and reflect real-world setups by residents. Radar frames were acquired continuously at a 10\, Hz rate, resulting in a total of approximately 43,000 frames. 

\subsection{FMCW Radar Signal Processing}
\label{subsec:preproc}
We use a 60 GHz FMCW radar to capture the complex baseband IQ frames. Each frame contains $N_c$ chirps, and each chirp has $N_{\mathrm{samples}}$. This linear chirp has a duration of $T_\mathrm{c}$, bandwidth $B$, and slope \cite{TI_FMCW_Training}:

\begin{equation}
\label{eq:chirp_slope}
S_w=B/T_\mathrm{c}. 
\end{equation}
After mixing (de-chirping) the received signal with the replica of the transmit signal (complex conjugate), we are able to obtain the intermediate-frequency (IF) beat signal. For a target at range $r$, the beat frequency is approximately \cite{TI_FMCW_Training}: 
\begin{equation}
    f_b= S_w\tau = \frac{S_w2r}{c} .
\end{equation}
where $c$ is the speed of light. We then estimate the range by applying a 1D Fast Fourier Transform (FFT) \cite{DUHAMEL1990259}, referred to as the Range-FFT, across the fast-time samples of each chirp. This results in range bins with a resolution defined by the radar bandwidth.
To capture motion information, for each range bin, we then apply a Doppler-FFT across the slow-time dimension (chirps), which results in a two-dimensional complex-valued matrix known as the Range-Doppler Map (RDM), $X_m[r, d]$. 
In our Single-Input and Multiple-Output  (SIMO) radar system with $M=3$ receivers, this process yields $M$ separate RDMs, $X_m[r, d]$, one for each receive antenna $m$. Since our goal is detect quasi-static post-fall floor occupancy (very weak motion), most of the useful signal is concentrated near zero Doppler, and clutter/multipath can dominate. To improve stability, we use lightweight preprocessing steps such as temporal smoothing window, magnitude normalization, and Moving Target Indicator filter (MTI)~\cite{8703820} before generating angle-dependent maps. For each bin $(r,d)$ in the RDM at frame k, the clutter estimation $C_k(r,d)$ is updated through~\cite{infineon_an141319}:
\begin{equation}
    C_k(r,d) = \alpha \cdot C_{k-1}(r,d) + (1-\alpha) \cdot RDM_k(r,d)
\end{equation}
\begin{equation}
    Y_k(r,d) = RDM_k(r,d) - C_k(r, d)
\end{equation}
where $\alpha$ is the forgetting factor ($\alpha = 0.01$ in our study). Importantly, the complex phase across antennas is preserved because it is needed for later spatial processing in \ref{subsec:dbf} and \ref{subsec:capon}. 
Although the RDM is a standard intermediate representation in many fall studies, our detection is performed on range-Azimuth (RA) feature maps. Our choice is motivated by the task that, after a fall, the subject is almost stationary. Therefore, in this case, its Doppler signatures are weak and unstable, while spatial localization on the floor is a more crucial indicator of post-fall occupancy.
\subsection{Vendor Digital Beamforming (DBF) Baseline}
\label{subsec:dbf}
We compare our proposed method against the vendor-provided digital beamforming (DBF) pipeline, which is a conventional approach for obtaining spatial information from a small antenna array.  Following the range–Doppler (RD) preprocessing and MTI described in Section~\ref {subsec:preproc}, the radar provides a per-range snapshot across the receiver antennas.
Particularly for our specific quasi-static application, we adapt the Infineon SDK code sample \texttt{range-angle-map.py} to our sensor configuration specified in Table \ref{tab:radar-hw}~\footnote{See Infineon’s DBF application notes for BGT60TR13C, we follow their geometry and weight definition. \cite{infineonAN155322,infineon_an141319}} .
Let $z_m(r,k)$ denote the complex RD value for receiver $m\!\in\!\{1,2,3\}$ at range bin $r$ and slow-time (Doppler) index $k$ after MTI. For a grid of look angles $(\theta,\phi)$, DBF computes the complex beam output by phasing and summing the channels; the corresponding power for frame $k$ is (\cite{infineon_an141319} Eq.~(6))
\begin{equation}
P_{\text{DBF}}(r,\theta,\phi;k) \;=\; \sum_{m=1}^{3} z_m(r,k)\,w_m(\theta,\phi).
\label{eq:dbf-power}
\end{equation}
The $w(\theta,\phi)$ is per-direction complex weights (steering) vectors and can be computed as (\cite{infineon_an141319} Eqs.~(4)–(5)):
\begin{equation}
\mathbf{w}(\theta,\phi) 
= 
\Bigl[
e^{j \tfrac{2\pi}{\lambda} \, d_x \cos\theta\cos\phi},\;
e^{j \tfrac{2\pi}{\lambda} \, d_y \sin\phi},\;
1
\Bigr].
\label{eq:weights}
\end{equation}
Following Infineon’s SDK, the 2D‐FFT DBF yields a 4‐D spectrum
$P_{\text{DBF}}(r,\theta,\phi;k)$ (range $r$, azimuth $\theta$, elevation $\phi$, frame $k$).
For visualization and downstream CA–CFAR, we obtain the RA map by
non–coherent integration across elevation $\phi$:
\begin{equation}
\label{eq:dbf-ra}
\mathrm{RA}_{\text{DBF}}(r,\theta;k)
~=~ \sum_{\phi\in\Phi} \bigl| P_{\text{DBF}}(r,\theta,\phi;k) \bigr|.
\end{equation}
\subsection{Proposed MVDR/Capon Range-Azimuth Processing}
\label{subsec:capon}
We keep the same FMCW front-end processing (Range/Doppler FFTs) but replace the spatial stage by a modified signal processing pipeline based on MVDR/Capon beamforming to improve robustness in cluttered, quasi-static conditions. For a fixed range bin $r$, we form a small set of Doppler snapshots by aggregating $N_D$ Doppler indices around a near-zero Doppler region since post-fall subjects are largely quasi-static and most useful signal remains close to zero velocity. Let ${D_1, D_2, ... D_{N_D}}$ denote the Doppler-bin indices, then the matrix $X_r$ stacks the complex antenna returns across receivers for those bins, where each column corresponds to one Doppler bin and each row corresponds to one receiver channel
\begin{equation}
  X_r \;=\;
  \begin{bmatrix}
    Y_1(r,D_1) & \cdots & Y_1(r,D_{N_D})\\[-1mm]
    \vdots     & \ddots & \vdots\\[-1mm]
    Y_{N_{\!Rx}}(r,D_1) & \cdots & Y_{N_{\!Rx}}(r,D_{N_D})
  \end{bmatrix}
  \;\in\; \mathbb{C}^{N_{\!Rx}\times N_D}.
  \label{eq:snapshot}
\end{equation}
This aggregation helps stabilize the covariance estimate when the target return is weak. The algorithm then estimates the Angle of Arrival (AoA) by solving an optimization problem that minimizes the output power $P_{\text{out}}$ while ensuring a distortionless response in the desired target direction: 
\begin{equation}
\label{eq:mvdr-opt}
\min_{\mathbf w} \; \Bigg(  P_{\text{out}} = \frac{1}{2} \mathbf w^{\mathrm H}\mathbf R_r\,\mathbf w \Bigg)
\quad \text{s.t.} \quad
\mathbf w^{\mathrm H}\mathbf a(\theta)=1,
\end{equation}
where $R_r$ is the sample spatial covariance at range $r$ that can estimated as following \eqref{eq:snapshot}:
\begin{equation}
  R_r \;=\; \frac{1}{N_D}\, X_r X_r^{\!H} \;\in\; \mathbb{C}^{N_{\!Rx}\times N_{\!Rx}}.
  \label{eq1:cov}
\end{equation}
and form its Moore–Penrose pseudoinverse $R_r^{-1}$ (no diagonal loading, $\delta=0$).
With azimuth steering vector:
\begin{equation}
  a(\theta) \;=\; \begin{bmatrix}1 \\ e^{-j\pi\sin\theta}\end{bmatrix},
  \label{eq:steer2el}
\end{equation}
the Capon power spectrum at range \(r\) and azimuth \(\theta\) is computed as
\begin{align}
  P_{\mathrm{Cap}}(r,\theta) \;&=\; \frac{1}{a(\theta)^{\!H}\,\widehat R_r^{-1}\,a(\theta)}.
  \label{eq1:capon}
\end{align}
Compared to the vendor DBF baseline, our proposed method based on MVDR/Capon algorithm introduces extra computation, which is mainly from estimating and inverting the spatial covariance matrix for each range bin. DBF can be implemented as a steering-vector projection per azimuth angle; its complexity is roughly $O(N_\theta N_{\mathrm{Rx}})$ per range bin, where $N_\theta$ is the number of azimuth grid points and $N_{Rx}$ is the number of receive channels. Our proposed method, on the other hand, requires forming $R_r$ using $N_D$ Doppler snapshots, inverting and evaluating the spectrum over azimuth. For our SIMO radar, where $N_{\mathrm{Rx}}$ is small, matrix inversion is inexpensive. Its complexity costs $O(N_\theta N_{\mathrm{Rx}}^2)$, where the main overhead is from $N_{\mathrm{Rx}}^2$ in spectrum evaluation.
\subsection{Cell-Averaging Constant False Alarm Rate}
\label{subsec:CA_CFAR}
\begin{figure}[htb] 
\centering 
\includegraphics[width=\linewidth]{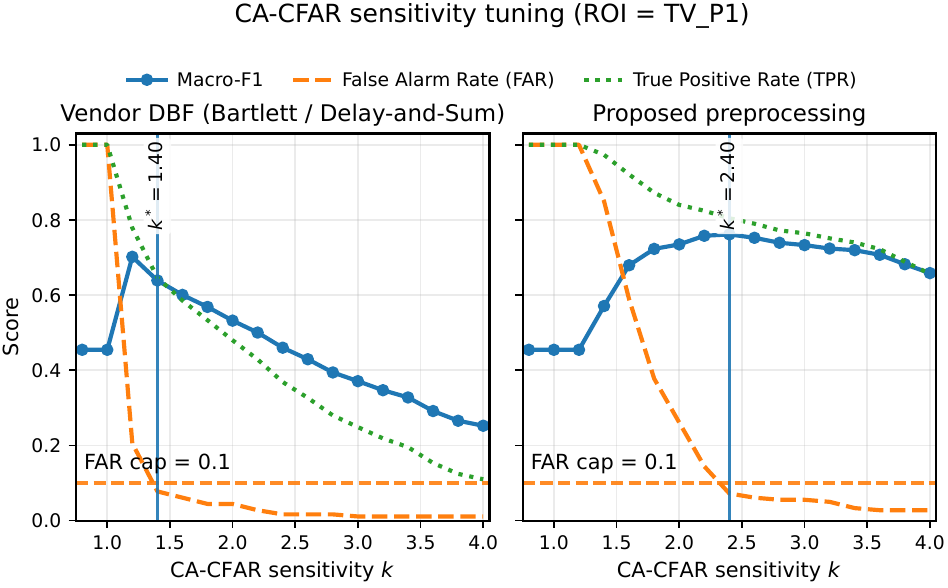}
\caption{CA-CFAR sensitivity tuning on Subject 1's dataset. Macro-F1, frame-level false positive rate (FPR), and true positive rate (TPR) are shown as the CA-CFAR scale factor $k$ is swept. The selected operating points $k$ satisfy the FPR cap (0.1) while maximizing Macro-F1 for each method.}
\label{fig:cfar_tuning_k_1x2}
\end{figure}

After generating RA maps using either DBF or MVDR/Capon method, we apply a 2D Cell-Averaging Constant False Alarm Rate (CA-CFAR) detector to identify energy target that correspond to a subject occupied floor region. Given each cell--under--test (CUT), CA-CFAR compares its power $|Z_{\mathrm{CUT}}|^2$ to an adaptive threshold derived from neighbouring reference cells~\cite{10289281,Miller2009FundamentalsOR}:
\begin{equation}
I_{\mathrm{det}}(\mathrm{CUT}) =
\begin{cases}
1, & |Z_\mathrm{CUT}|^2 > k \frac{1}{N_\text{train}} \sum_{(i,j)\in\mathcal{T}(r,c)}Y(i,j),\\[1ex]
0, & \text{otherwise},
\end{cases}
\label{eq:cfar_basic}
\end{equation}
where $\mathcal{T}(r,c)$ be the set of training cells, $N_\text{train}$ is the number of training samples, and $k$ controls sensitivity. In our work, we tune the CA-CFAR sensitivity parameter $k$ using Subject 1’s tuning subset by sweeping across a practical range of values and selecting an operating point that controls false alarms while maintaining high detection performance. We treat each radar frame as a binary decision: a positive frame means the subject is present on the floor (post-fall occupancy within the ROI), while a negative frame means the room is empty. Thus, we defined the frame-level false positive rate (FPR)--also referred to as false alarm rate (FAR) in Fig.~\ref{fig:cfar_tuning_k_1x2}--as the fraction of negative frames that produce at least one CFAR detection inside the region-of-interest (ROI),
\begin{equation}
    FPR = \frac{FP}{(FP + TN)},
\end{equation}
where FP counts empty-room frames. 
We emphasize that this FPR cap is a frame-level constraint used for detector tuning only. In deployment, alarms are not triggered from a single frame. Instead, a temporal integration rule are applied to trigger only if the ROI is detected in a large fraction of frames over a short window, such as 20 seconds.
We choose an FPR cap of 0.1 ($\text{FPR} \leq 0.1$) 
we select k by maximizing Macro-F1, which balances both detection (TPR) and missed frames (FNR) on the tuning subset. This produces a working operating point that balances both missed detections caused by weak quasi-static signatures and false alarms caused by clutter leakage.

Fig.~\ref{fig:cfar_tuning_k_1x2} shows the sweep for both the vendor DBF baseline and our proposed preprocessing. It can be seen from the figure that increasing k reduces false alarms but can also suppress weak quasi-static returns, which affects both Macro-F1 and TPR. Under the same FPR cap, the selected operating points differ: the DBF baseline reaches its chosen point around k = 1.4, while our proposed preprocessing operates at a higher value k = 2.4. Importantly, when we evaluate empty-room data across all subjects and views, the DBF pipeline achieves about 99\% true-negative rate (TNR), while the proposed pipeline achieves 100\% TNR, corresponding to approximately 1\% and 0\% FPR, respectively.
\begin{table*}[t]
\caption{Participant-level frame-positive rate by viewpoint and location (higher is better). Best values per view and location for each subject are shown in bold.}
\label{tab:trial_matrix}
\setlength{\tabcolsep}{6.0pt}
\renewcommand{\arraystretch}{0.98}
\scriptsize
\resizebox{\textwidth}{!}{
\begin{tabular}{@{} l l r r r r r r r r r r r r r r @{}}
\toprule
\multirow{2}{*}{\textbf{View}} & \multirow{2}{*}{\textbf{Loc.}}
& \multicolumn{2}{c}{\textbf{Subject 1}} & \multicolumn{2}{c}{\textbf{Subject 2}} & \multicolumn{2}{c}{\textbf{Subject 3}} & \multicolumn{2}{c}{\textbf{Subject 4}} & \multicolumn{2}{c}{\textbf{Subject 5}} & \multicolumn{2}{c}{\textbf{Subject 6}} & \multicolumn{2}{c}{\textbf{Subject 7}} \\
& & \textbf{DBF} & \textbf{Prop.} & \textbf{DBF} & \textbf{Prop.} & \textbf{DBF} & \textbf{Prop.} & \textbf{DBF} & \textbf{Prop.} & \textbf{DBF} & \textbf{Prop.} & \textbf{DBF} & \textbf{Prop.} & \textbf{DBF} & \textbf{Prop.} \\
\cmidrule(lr){3-4} \cmidrule(lr){5-6} \cmidrule(lr){7-8} \cmidrule(lr){9-10} \cmidrule(lr){11-12} \cmidrule(lr){13-14} \cmidrule(lr){15-16}
\midrule
\multirow{6}{*}{TV} & P1 & 0.35 & \textbf{0.57} & 0.78 & \textbf{0.92} & 0.72 & \textbf{0.93} & 0.59 & \textbf{0.90} & 0.75 & \textbf{0.88} & 0.88 & \textbf{0.93} & 0.54 & \textbf{0.66} \\
 & P2 & 0.98 & \textbf{0.99} & 0.98 & \textbf{0.99} & 0.70 & \textbf{0.72} & 0.98 & \textbf{0.99} & 0.58 & \textbf{0.70} & 0.97 & \textbf{0.98} & 0.97 & \textbf{0.99} \\
 & P3 & 0.97 & \textbf{0.99} & 0.99 & \textbf{1.00} & 0.77 & \textbf{0.91} & 1.00 & 1.00 & 1.00 & 1.00 & 0.99 & \textbf{1.00} & 0.96 & \textbf{0.99} \\
 & P4 & 0.27 & \textbf{0.40} & 0.77 & \textbf{0.94} & 0.99 & \textbf{1.00} & 0.98 & \textbf{1.00} & 0.97 & \textbf{1.00} & 0.97 & \textbf{1.00} & 0.99 & 0.99 \\
 & P5 & 0.63 & \textbf{0.94} & 0.97 & \textbf{1.00} & 0.33 & \textbf{0.67} & 0.88 & \textbf{0.94} & 0.47 & \textbf{0.78} & 0.87 & \textbf{0.98} & 0.97 & \textbf{0.98} \\
 & \textit{Mean} & 0.64 & \textbf{0.78} & 0.90 & \textbf{0.97} & 0.70 & \textbf{0.85} & 0.89 & \textbf{0.97} & 0.75 & \textbf{0.87} & 0.94 & \textbf{0.98} & 0.89 & \textbf{0.92} \\
\midrule
\multirow{6}{*}{Window} & P1 & 0.63 & \textbf{0.90} & 0.91 & \textbf{0.99} & 0.88 & \textbf{0.99} & 0.81 & \textbf{0.97} & 0.81 & \textbf{0.98} & 0.92 & \textbf{0.97} & 0.63 & \textbf{0.94} \\
 & P2 & 0.48 & \textbf{0.73} & 0.87 & \textbf{0.99} & 0.48 & \textbf{0.69} & 0.62 & \textbf{0.85} & 0.44 & \textbf{0.71} & 0.84 & \textbf{0.92} & 0.54 & \textbf{0.74} \\
 & P3 & 0.99 & 0.99 & 1.00 & 1.00 & 0.98 & \textbf{0.99} & 0.87 & \textbf{0.92} & 0.96 & \textbf{0.98} & 0.95 & \textbf{0.97} & 0.95 & \textbf{0.98} \\
 & P4 & 1.00 & 1.00 & 1.00 & 1.00 & 0.98 & \textbf{1.00} & \textbf{0.96} & 0.93 & 1.00 & 1.00 & 0.99 & 0.99 & 0.94 & \textbf{0.95} \\
 & P5 & 0.70 & \textbf{0.74} & 1.00 & 1.00 & 0.88 & \textbf{0.94} & 1.00 & 1.00 & 1.00 & 1.00 & 0.97 & \textbf{0.98} & 0.99 & \textbf{1.00} \\
 & \textit{Mean} & 0.76 & \textbf{0.87} & 0.96 & \textbf{1.00} & 0.84 & \textbf{0.92} & 0.85 & \textbf{0.93} & 0.84 & \textbf{0.93} & 0.93 & \textbf{0.97} & 0.81 & \textbf{0.92} \\
\midrule
\multicolumn{2}{l}{\textbf{Overall}} & 0.70 & \textbf{0.83} & 0.93 & \textbf{0.98} & 0.77 & \textbf{0.88} & 0.87 & \textbf{0.95} & 0.80 & \textbf{0.90} & 0.94 & \textbf{0.97} & 0.85 & \textbf{0.92} \\
\bottomrule
\end{tabular}%
}
\end{table*}
\section{Results}
\begin{figure}[htb] 
\centering
\includegraphics[width=\linewidth]{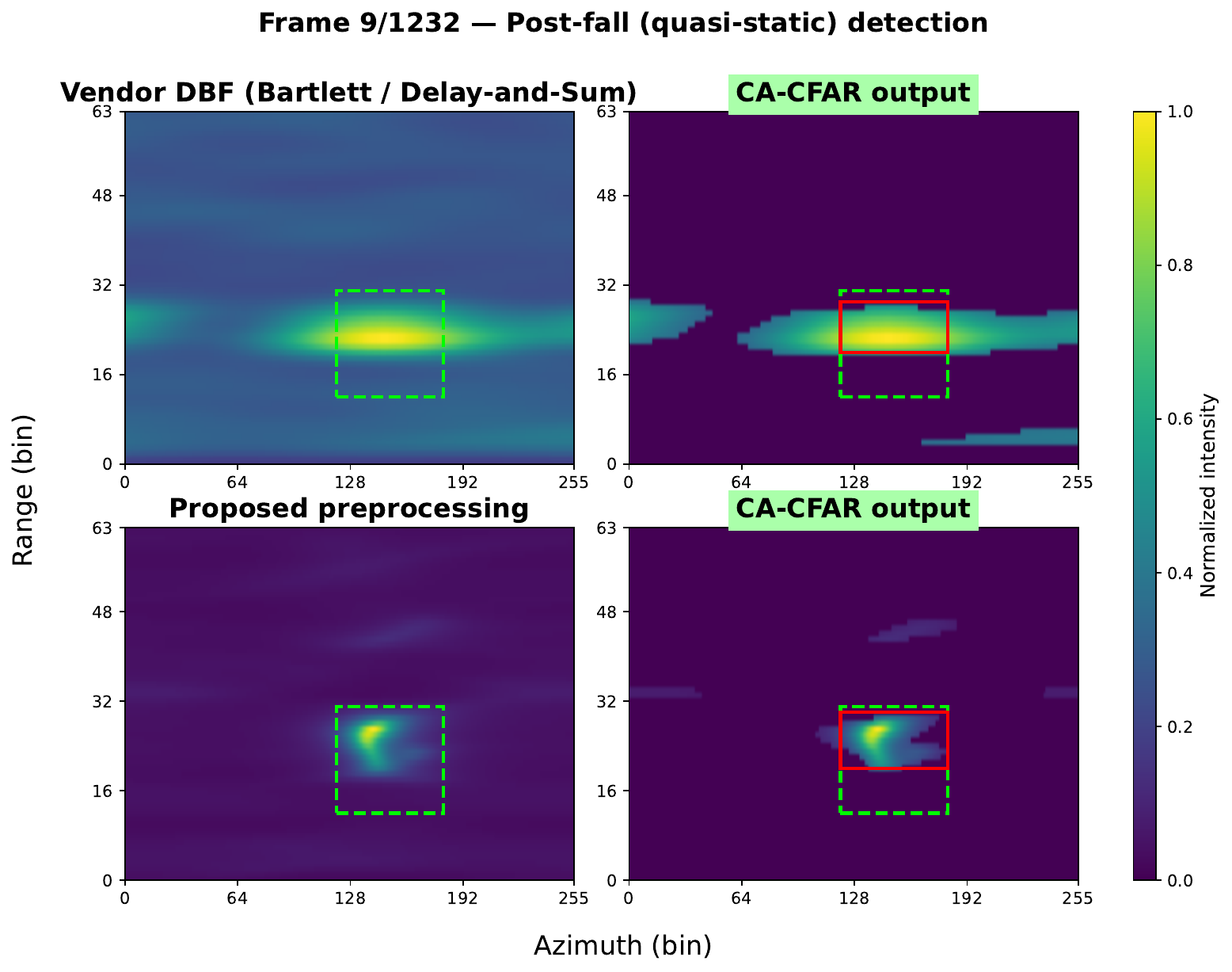}
\caption{Qualitative comparison of DBF and our proposed detectors on frame 9 for subject 1 lying on the floor at position 1 (window view), which both of them produce detections within the ground-truth bounding box.}
\label{fig:both_detect}
\end{figure}
\begin{figure}[htb] 
\centering 
\includegraphics[width=\linewidth]{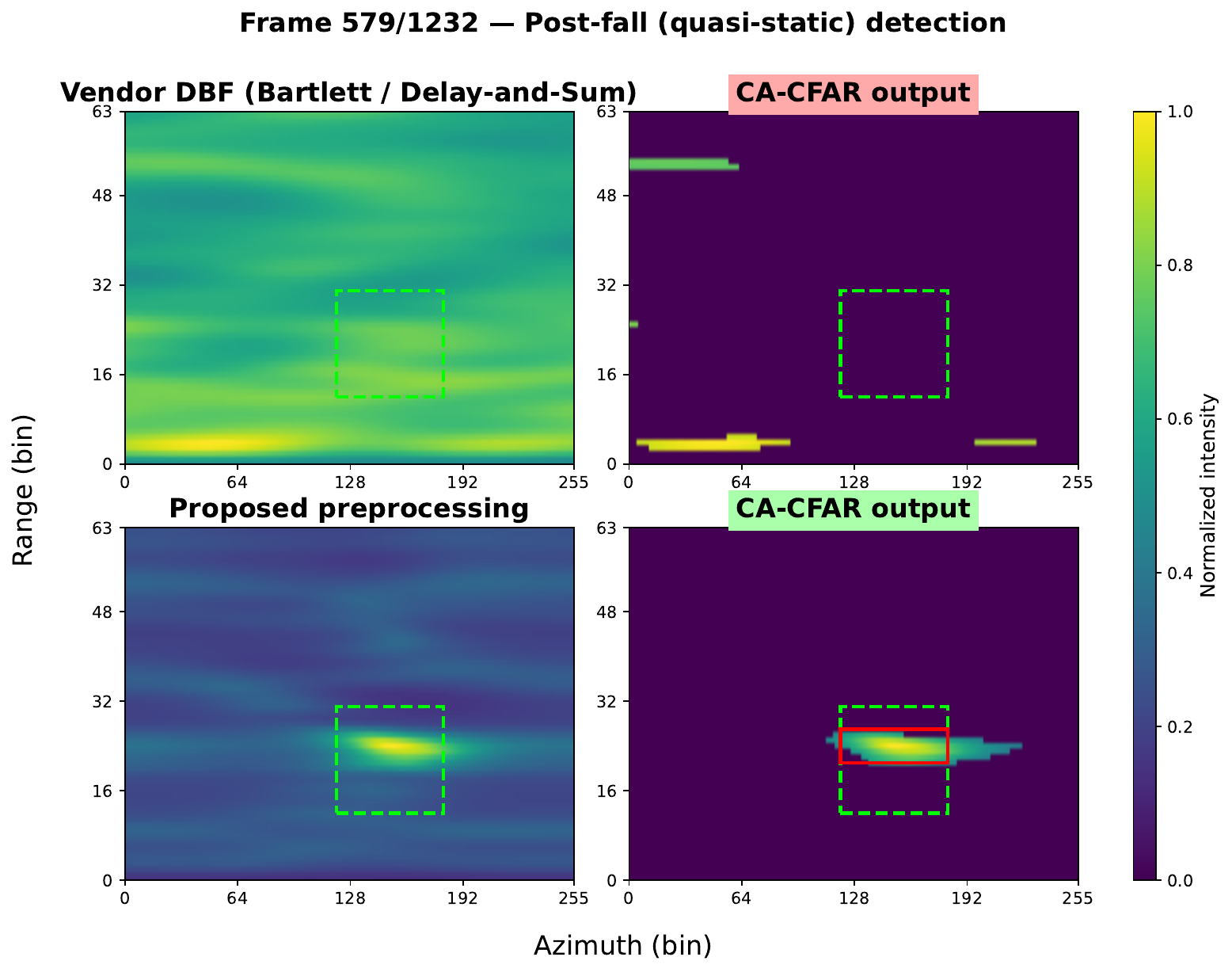}
\caption{Qualitative comparison of DBF and our proposed detectors on frame 579 for subject 1 lying on the floor at position 1 (window view). Only our proposed method can detect within the ground-truth bounding box.}
\label{fig:one_detect}
\end{figure}
After fixing k for each pipeline, we keep it unchanged for all remaining subjects, viewpoints, and locations. Figs.~\ref{fig:both_detect} and~\ref{fig:one_detect} qualitatively compare the detection results between the vendor DBF and the proposed detector on two representative frames of subject 1 lying on the floor at position 1. 
In both figures, the subject’s approximated post-fall location is annotated by a green dashed bounding box in the range–azimuth (RA) map. 
For each location (for example, TV view and position 1), we determine a center point $(r_0,\theta_0)$ from the known floor placement during data collection and a one-time calibration of the radar coordinate system. We then use a fixed-size box around this center, $[r_0\pm\Delta r,\ \theta_0\pm\Delta\theta]$ with $\Delta r$ and $\Delta\theta$ kept constant across frames and subjects for a given viewpoint/location to ensure consistent scoring. The chosen values are wide enough to capture the expected spread of the subject’s return across nearby range and azimuth bins (due to posture variation, small placement shifts, and limited angular resolution), but still narrow enough to avoid counting unrelated clutter regions as true detections. We mark a frame as a hit if one remaining CA-CFAR detection (after thresholding and suppression) overlaps the annotated ground–truth bounding box. A green highlight indicates a successful detection, while a red highlight signifies a miss. Fig.~\ref{fig:both_detect} shows a typical case where both methods successfully localize the subject within the ground-truth region, resulting in both results being counted, and the labels of the CA-CFAR results are highlighted in green. However, Fig.~\ref{fig:one_detect} shows a more challenging frame where strong background clutter can affect DBF's result, CA-CFAR fails to localize the target and leaves with only clutter. Our proposed method, on the other hand, demonstrates superior stability under severe noise and succeeds in detecting the subject within the ground-truth bound. We quantify performance using the frame-positive rate (fraction of frames in a trial that are counted as hits).
\begin{figure}[htb] 
\centering 
\includegraphics[width=\linewidth]{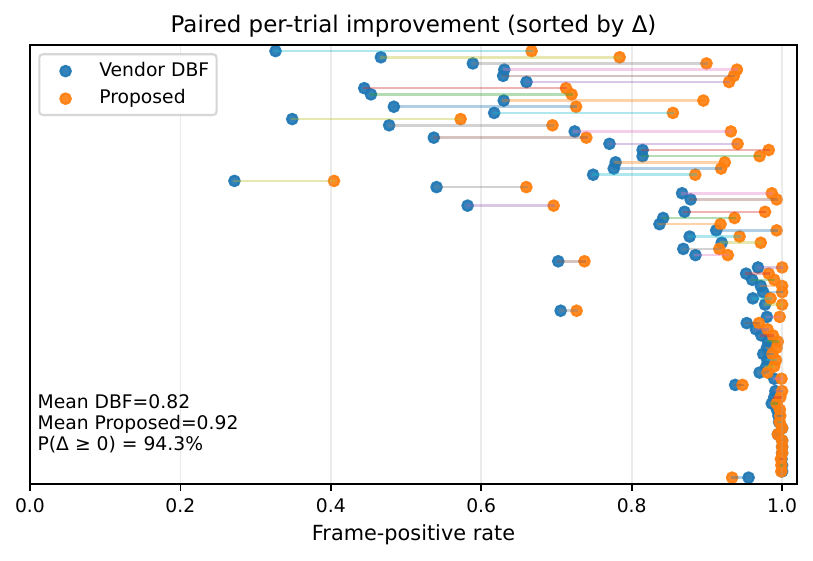}
\caption{Paired per-trial comparison of frame-positive rate. Each point corresponds to one trial (subject, view, and location). Trials are sorted by improvement $\Delta$.}
\label{fig:dumbbell_sorted_overall}
\end{figure}

Table \ref{tab:trial_matrix} breaks down the accuracy detection by viewpoint (TV-side vs Window-side) and floor locations (P1-P5) for seven subjects. Many trials are already near 100\% accuracy for both methods, but the largest gains appear in harder cases where DBF struggles. For example, Subject~1's accuracy at TV--P1 increases from 0.35 to 0.57, and Subject~1's accuracy at TV--P5 increases from 0.63 to 0.94. The worst DBF case (0.27 at TV--P4 for subject~1) also improves to 0.40. These improvements suggest that the proposed method is especially helpful when the post-fall signature is weak, partially blocked, or mixed with static clutter. We note that even with the improvement, TV-P4 for Subject 1 remains a challenging condition, where many frames are missed due to strong clutter and geometry (e.g., reflections from nearby furniture or partial blockage). As mentioned above in~\ref{subsec:CA_CFAR}, in a real-life safety-critical deployment workflow, post-fall detection is not a one-frame decision, but temporal integration can turn a moderate per-frame hit rate into a reliable alarm over a short window. For example, if at least 80\% of frames are detected over 20 seconds, an alert is triggered.
\begin{figure}[htb] 
\centering 
\includegraphics[width=\linewidth]{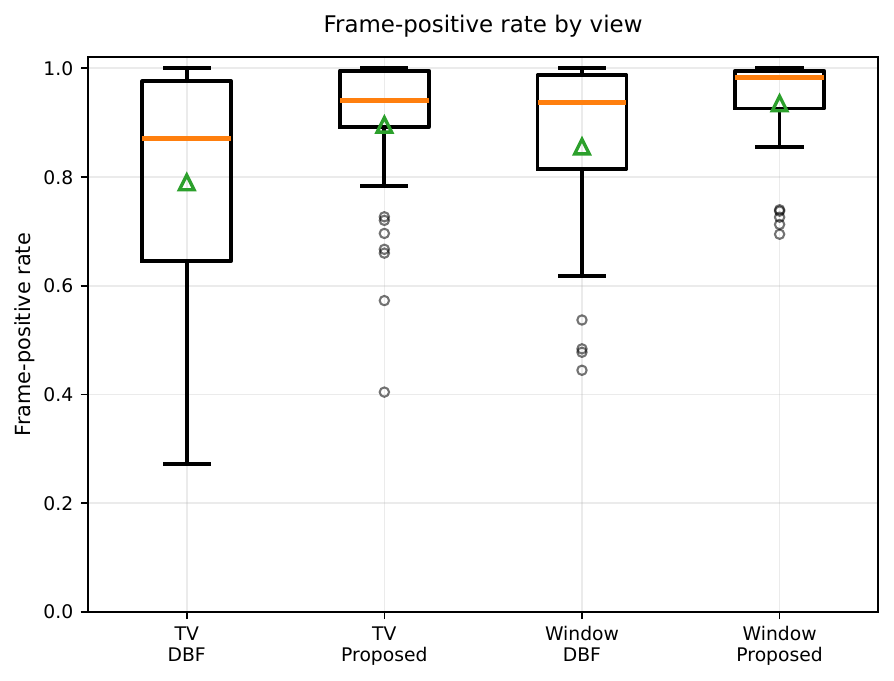}
\caption{Frame-positive rate by viewpoint. Boxplots compare vendor DBF and proposed method for TV-side and Window-side views.}
\label{fig:boxplot_by_view}
\end{figure}

Fig.~\ref{fig:dumbbell_sorted_overall} plots paired trial-level frame-positive rates for DBF and the proposed method, sorted by the improvement $\Delta$. Across all paired trials, the proposed method increases the mean frame-positive rate from 0.82 (DBF) to 0.92 (Proposed), and improves or matches DBF in 94.3\% of trials. This indicates that the gain is not limited to a small subset of cases, but is consistent across subjects, viewpoints, and post-fall locations.
\begin{figure}[t] 
\centering 
\includegraphics[width=\linewidth]{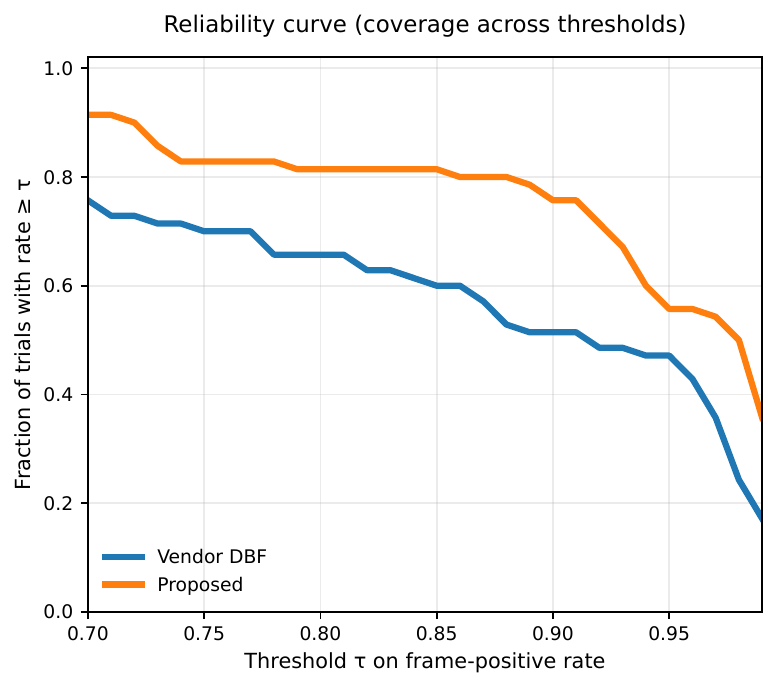}
\caption{Reliability curve (coverage across thresholds). The y-axis shows the fraction of trials with frame-positive rate $\ge \tau$.}
\label{fig:coverage_curve}
\end{figure}
To summarize the distribution shift, Fig.~\ref{fig:boxplot_by_view} reports boxplots of frame-positive rates grouped by viewpoint and method. For both TV-side and Window-side views, the proposed method shifts the distribution upward compared to the DBF baseline. This outlier improvement matters in a clinical monitoring context.
In a clinical monitoring context, reducing these low-performing trials is crucial since those are the situations most likely to produce missed post-fall detection. And finally, to better quantify the robustness at different accuracy requirements, we report the reliability curve in Fig.~\ref{fig:coverage_curve}.
We measure the fraction of trials whose frame-positive rate is at least $\tau$. This is crucial in a clinical setting since the system is expected to remain reliable and sustained rather than occasional correct detection.
As shown in Fig.~\ref{fig:coverage_curve}, the proposed method outperforms the DBF baseline across various thresholds, meaning it maintains higher coverage even as the success criterion becomes stricter. This suggests our method not only improves the mean score, but also in the worst-case reliability, which is crucial for reducing missed post-fall detections in difficult layouts and subject positions.

\label{sec:results}
\section{Conclusion}
In conclusion, this paper compared a proposed preprocessing pipeline against a vendor DBF baseline for quasi-static post-fall floor-occupancy detection using a low-cost 60 GHz FMCW radar in realistic indoor rooms with furniture and strong multipath. Instead of focusing on fall-event classification, we target the less well-addressed and more challenging clinical need of reliably detecting the post-fall state, where the subject remains mostly quasi-static on the floor, and floor reflections and clutter can dominate the radar return signals. Our proposed method, based on Capon/MVDR beamforming, provides more stable detections and higher reliability across subjects and room locations. Across our evaluation, the proposed approach achieves a mean frame-positive rate of about 0.92 and maintains higher coverage under stricter success thresholds compared to the vendor DBF baseline.

Future work will expand the pipeline toward a more complete monitoring workflow that combines fall and post-fall detection, including better use of range Doppler and angle data representation across time. Another important direction is post-fall vital-sign monitoring, which could provide additional clinical value while waiting for caregiver response after an incident.

\label{sec:conclusion}
\section*{Acknowledgment}
The authors would like to give thanks to ElephasCare, Schlegel-UW Research Institute for Aging, and MITACS for supporting this study.

\bibliographystyle{IEEEtran}
\bibliography{refs}

\end{document}